\documentclass[aps,prl,floatfix,twocolumn]{revtex4-2}
\usepackage{graphicx}
\usepackage{bm,amsmath}
\usepackage{dcolumn}
\usepackage{amsfonts,amssymb}
\usepackage{chemfig}
\usepackage{subfigure}
\usepackage[colorlinks=true,linkcolor=blue,citecolor=blue,urlcolor=blue]{hyperref}

\begin{document}

\title{Spinons and Spin-Charge Separation at the Deconfined Quantum Critical Point}

\author{Sibin Yang}
\affiliation{Department of Physics, Boston University, 590 Commonwealth Avenue, Boston, Massachusetts 02215, USA}

\author{Anders W. Sandvik}
\email{sandvik@bu.edu}
\affiliation{Department of Physics, Boston University, 590 Commonwealth Avenue, Boston, Massachusetts 02215, USA}

\date{\today}

\begin{abstract}
Using quantum Monte Carlo and numerical analytic continuation methods, we study the dynamic spin structure factor and the single-hole 
spectral function of a two-dimensional quantum magnet ($J$-$Q$ model) at its quantum phase transition separating N\'eel antiferromagnetic 
and spontaneously dimerized ground states. At this putative deconfined quantum-critical point, we find a broad continuum of spinon 
excitations that can be accounted for by the fermionic $\pi$-flux state; a known mean-field model for deconfined 
quantum criticality. We find that the best description of the two-spinon continuum is with a version of the model with a $2\times 2$ 
unit cell, reflecting non-trivial mutual statistics of spinons and anti-spinons. The single-hole spectral function can be described 
by the same spinon dispersion relation and an independently propagating holon. Thus, the system exhibits spin-charge separation 
and will likely evolve into an extended holon metal phase at finite doping.
\end{abstract}

\maketitle

The proposal of a "beyond Landau" deconfined quantum critical point (DQCP) \cite{senthil04a} separating the N\'eel antiferromagnetic (AFM) ground state of 
two-dimensional (2D) $S=1/2$ quantum magnets from a spontaneously dimerized valence-bond solid (VBS) is still under intense debate. To test predictions 
of various field-theoretical scenarios \cite{read89,read90,murthy90,senthil04b,levin04,hermele04,hermele05,tanaka05,senthil06,dyer15} that typically are reliable 
only in certain limits, e.g.,  with SU($N$) at large $N$, the AFM--VBS transition has been studied extensively using Monte Carlo simulations of lattice models
\cite{sandvik07,jiang08,melko08,kaul08,kuklov08,lou09,sandvik10a,kaul12,chen13,block13,nahum15a,nahum15b,shao16,suwa16,sreejith19,zhao19,serna19,zhao20,sandvik20,takahashi20,wang21,zhao22,sato23,takahashi24,demidio24,song25}
as well as modern numerical conformal field theory (CFT) approaches in 2+1 dimensions 
\cite{nakayama16,li19,poland19,chester24,zhou24,chen24}. 
Two competing scenarios have emerged from these studies: 
(i) a DQCP requiring a complex CFT description, only realized as a weak first-order transition in conventional lattice models
\cite{wang17,gorbenko18a,gorbenko18b,ma20,nahum20,he21,zhou24,chen24,song25}, 
and (ii) the weak first-order transitions that indeed are commonly found in Monte Carlo studies of SU($2$) lattice models 
reflect a nearby multicritical point that exists within these models \cite{takahashi20,yang22,chester24}, but only beyond the parameter regime for which 
a positive definite sampling space can be constructed \cite{takahashi24}. The putative multicritical point would then likely also be the tip 
of a gapless spin liquid phase between the AFM and VBS phases in an extended parameter space \cite{schackleton21,yang22,liu22,liu24,chen24}.

Regardless of these unresolved scenarios, the very weak first-order transitions studied on the square lattice are sufficiently 
close to the DQCP for critical fluctuations to be manifested up to very large length scales \cite{takahashi24}. Moreover, spinon deconfinement
has been observed in the dynamic spin structure factor, with a dispersion relation  
in remarkably good agreement with that of the $\pi$-flux phase of fermions on the square lattice, 
which is one  of the candidate mean-field descriptions of the DQCP \cite{ma18}. Another important aspect of deconfinement has  
until now not been addressed with lattice calculations: the fractionalization of an injected hole 
in the host quantum magnet. Putative spin-charge separation has consequences also at finite doping, where a
"holon metal" that is unstable to superconductivity has been proposed based on an extension of the DQCP scenario \cite{kaul07}. 

Upon doping, the fermionic aspects of spin models become manifested and quantum Monte Carlo (QMC)
studies are hampered by mixed signs of the sampling weights \cite{loh90}. Results can then not be obtained on the same large 
scales as with pure spin models. The single-hole spectral function corresponding to injection of a single 
hole can be sampled, however, and is pursued here. The Green's function is computed in imaginary time using QMC
simulations with the approximation-free Stochastic Series Expansion (SSE) method \cite{sandvik99,sandvik10} and is subsequently analytically continued 
to real frequency using a method capable of resolving sharp features \cite{shao23}, e.g., edges expected with fractionalized excitations. 

We present two related findings: The dynamic spin structure factor of the host quantum magnet can be well accounted for 
by the $\pi$-flux model. However, in contrast to previous work \cite{ma18},  we argue that a $2\times 2$
unit cell for the phase pattern should be used, leading to two degenerate filled and unfilled bands in a Brillouin zone
(BZ) $1/4$ of the size of the BZ of the simple square lattice. The dispersion relation is the same as the model with
$2\times 1$ unit cell (single filled and unfilled bands) used in  Ref.~\cite{ma18}. However, the support of the two-spinon
continuum in frequency $\omega$ is different for some momenta ${\bf k}$ in the unfolded BZ of the simple square lattice,
with the $2\times 2$ model providing a better match with the numerical results. A consistent description of the single-hole spectral 
function with fractionalized spinon and holon excitations is then also obtained. The system exhibiting spin-charge separation 
in addition to spinon deconfinement, is a candidate for the holon metal state \cite{kaul07} of the doped DQCP.

{\it Models and Methods.}---We study the ground state of a variant of the $S=1/2$ $J$-$Q$ model, defined
by the Hamiltonian \cite{lou09,takahashi24}
\begin{equation}
H_{JQ}=-J\sum_{\langle ij\rangle}P_{ij}-Q\sum_{\langle ijklmn\rangle}P_{ij}P_{kl}P_{mn},
\end{equation}
where $P_{ij}=1/4-{\bf S}_i \cdot {\bf S}_j$ is the singlet projector on nearest-neighbor sites $i,j$ on periodic $L\times L$
lattices. The $J$ term is the standard Heisenberg model and the $Q$ term, with the three projectors stacked on $3\times 2$ 
and $2\times 3$ lattice cells, favors columnar-correlated singlets, such that a VBS state forms above the transition point 
$(J/Q)_c \approx 0.6667$. While a $Q$ term with only two projectors \cite{sandvik07} also hosts a VBS state for large $Q/J$, 
more robust VBS order is achieved with three projectors. In an accompanying longer paper \cite{yang25a}, 
we also discuss the excitations inside the AFM and VBS phases.
 
In the presence of an injected hole, standard nearest-neighbor hopping is added,
\begin{equation}
H=H_{JQ}-t\sum_{\langle ij\rangle}\sum_{\sigma=\uparrow,\downarrow}(c^\dagger_{\sigma,i}c_{\sigma,j}+{\rm h.c.}),
\end{equation}
to describe the mobility of manifestly fermionic $S=1/2$ particles created and destroyed by the operators $c^\dagger_{\sigma,i}$ and 
$c_{\sigma,i}$. In sampling the hole paths in the SSE generated space-time background, we utilize a canonical
transformation by Angelucci \cite{angel95} that has previously been used with world-line QMC simulations of the 2D $t$-$J$ model
(our model with $Q=0$) \cite{brunner00}. Adaptation to SSE is described in the context of spin chains 
in Ref.~\cite{yang25b} and the extensions to the 2D square lattice  are straight-forward.

In addition to studying the single-hole spectral function $A({\bf k},\omega)$, we also revisit the dynamic spin structure factor $S({\bf k},\omega)$
of the host quantum magnet. It was studied with a spin-anisotropic variant of the $J$-$Q$ model in Ref.~\cite{ma18}, which found that the lower edge 
of the continuum (the dispersion relation of a spinon) is well described by a mean-field ansatz for four-flavor quantum 
electrodynamics in 2+1 dimensions; the $\pi$-flux state of fermions in a half-filled square lattice. As mentioned, 
we will discuss a different variant of the $\pi$-flux state, which also has consequences for spin-charge separation.

{\it Spin Dynamics.}---The dynamic spin structure factor is given in the Lehman representation by 
\begin{equation}
S({\bf k},\omega)=\sum_{n} |\langle n|S^z_{\bf k}|0\rangle|^2 \delta(\omega+E_0-E_n),
\end{equation}
for the quantum magnet with eigenstates $|n\rangle$ and energies $E_n$. Here we assume that the system is in its ground state 
$|0\rangle$, to which our QMC simulations converge at temperatures $T \ll L^{-1}$. $S^z_{\bf k}$ is the Fourier transform of the 
real-space spin operator, whose correlation function $G_S({\bf r},\tau)=\langle S^z_{\bf r}(\tau)S^z_0(0)\rangle$ at distance ${\bf r}$ and imaginary-time 
separation $\tau$ we compute with SSE QMC simulations and continue to real frequency using the most recent variant of stochastic analytic
continuation (SAC) \cite{shao23}. In order to resolve the sharp edges expected with fractionalized excitations, we apply 
constraints corresponding to a power-law singular lower edge of the spectrum, at a location that the method finds, followed by a 
continuum that is sampled according to procedures explained in previous works \cite{shao23,yang25b,yang25c}.

\begin{figure}[t!]
\includegraphics[width=8.4cm]{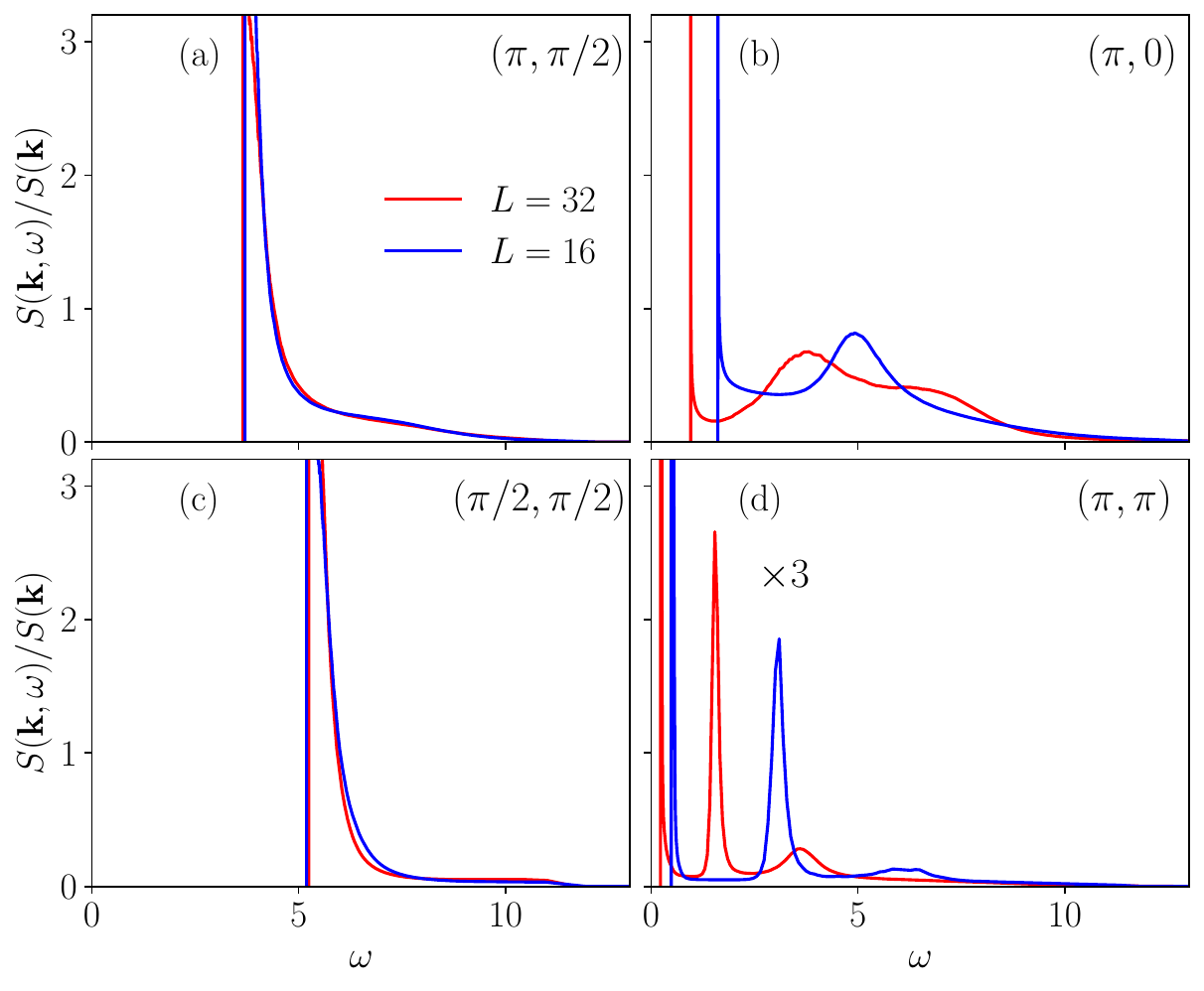}
\caption{Dynamic spin structure factor $S({\bf k},\omega)$ of the $J$-$Q$ model at the AFM--VBS transition. Results are shown for 
system sizes $L=16$ and $L=32$ at four different momenta $(k_x,k_y)$. The static structure factor $S({\bf k})$ is used
to normalize all spectra to unity; for $L=32$, $S(\pi,\pi/2)=1.142$, $S(\pi,0)=0.516$, $S(\pi/2,\pi/2)=0.564$ and $S(\pi,\pi)=13.4$.
The ${\bf k}=(\pi,\pi)$ results are multiplied by $3$.}
\label{fig1}
\end{figure}

At the critical point ($Q=1$, $J=0.6667$ \cite{lou09,takahashi24}) and lattice sizes $L=16$ and $32$,
Fig.~\ref{fig1} shows examples of $S({\bf k},\omega)$ at high-symmetry points in the BZ. The finite-size effects are only minor 
for the high-energy momenta $(\pi,\pi/2)$ and $(\pi/2,\pi/2)$, and the $L=32$ results should be very close
to size converged. In contrast, for $(\pi,0)$ and $(\pi,\pi)$, both of which are expected to be gapless points at the DQCP 
\cite{suwa16,ma18}, the edge moves down roughly as $L^{-1}$, which is expected with the dynamic
exponent $z=1$. The three peaks at $(\pi,\pi)$ likely only reflect the leading contributions to a finite-size multi-peak
structure that slowly develops into a narrow continuum above the edge for $L \to \infty$. The broad continua in Figs.~\ref{fig1}(a) and \ref{fig1}(c)
indicate fractionalized excitations; two spinons whose individual momenta ${\bf k}_1$ and ${\bf k}_2$ take a range of values for a
total momentum ${\bf k}={\bf k}_{1}+{\bf k}_2$. 

\begin{figure}[t!]
\includegraphics[width=8.4cm]{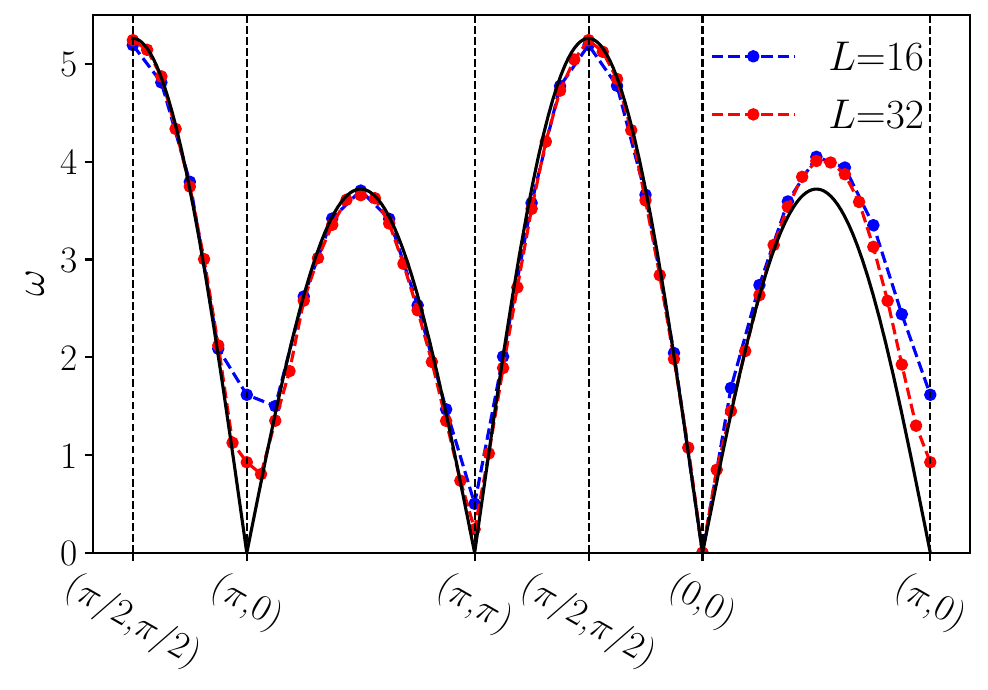}
\caption{Location of the edge in $S({\bf k},\omega$) versus the momentum along a path in the BZ, based on results such
as those in Fig.~\ref{fig1} for $L=16$ and $L=32$. The black curve shows a fit to the $\pi$-flux dispersion relation
Eq.~(\ref{epsilon_s}) with velocity $v=3.73$.}
\label{fig2}
\end{figure}

\begin{figure}[t!]
\includegraphics[width=8.7cm]{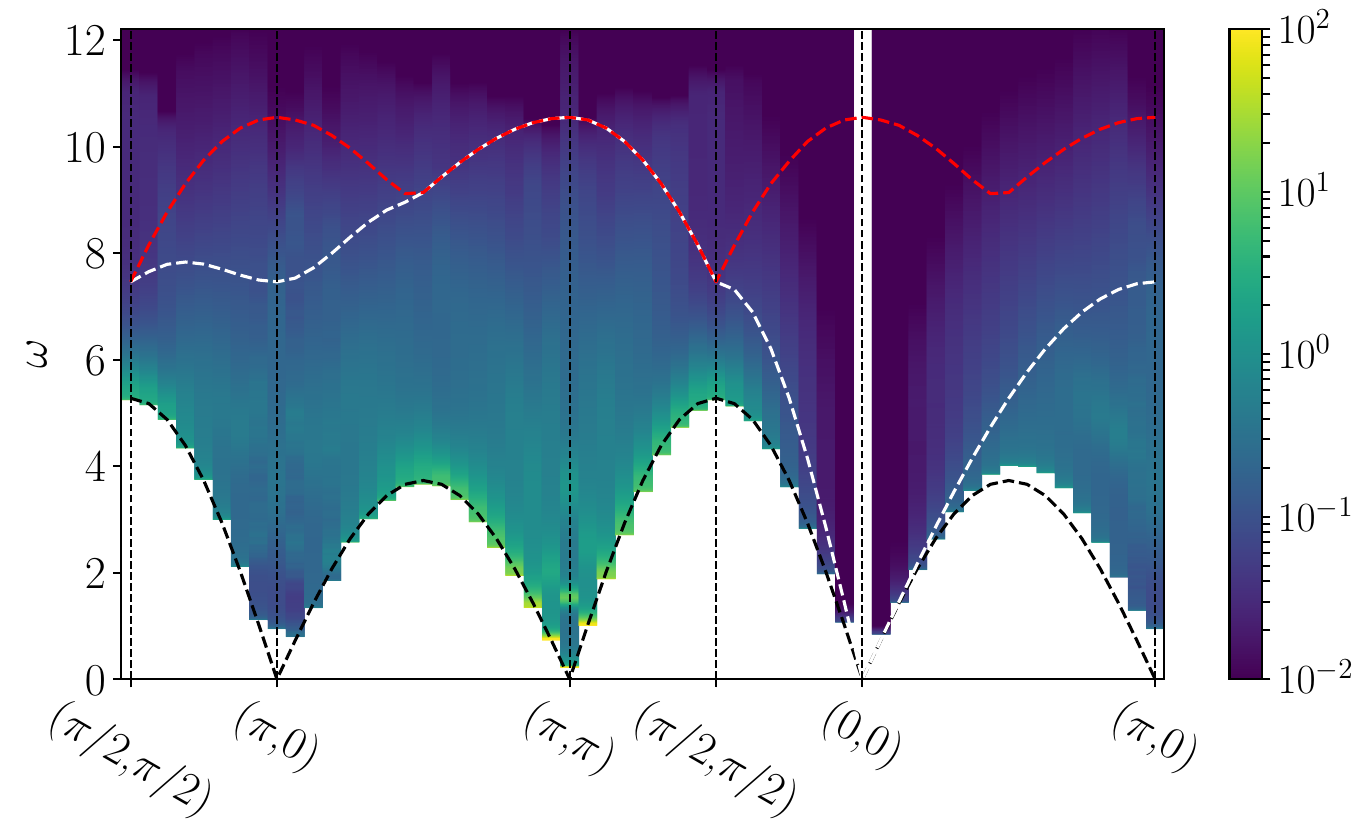}
\caption{Heat map of $S({\bf k},\omega)$ for $L=32$. There is no spectral weight in the white area, while weight below $10^{-2}$
above the edge is shown as as black. The black dashed curve is the same fit to the edge as in Fig.~\ref{fig2}. The red dashed curve shows
the upper bound of the two-spinon continuum predicted from this dispersion relation with the diamond shaped BZ
of size $1/2$ of the original square lattice BZ. The white dashed curve corresponds to the further reduced BZ with ${\bf k} \in [0,\pi]^2$.}
\label{fig3}
\end{figure}

Figure \ref{fig2} shows the dispersion relation $\omega_{\bf k}$ (the edge location) along a standard path in the BZ. 
In accord with Fig.~\ref{fig1}, significant finite-size effects are observed mainly at low energies, but the
convergence is slower also at higher energies on the segment between $(\pi/2,0)$ and $(\pi,0)$. In most of the BZ, the well 
converged $L=32$ energies can be accounted for by the dispersion relation of the excitations of the half-filled $\pi$-flux
model;
\begin{equation}\label{epsilon_s}
\epsilon_S({\bf k}) = v\sqrt{\sin^2(k_x)+\sin^2(k_y)},
\end{equation}
where $v$ is a model dependent velocity. With the best overall fit to the converged energies for $\omega \ge 1$, there
are significant deviations only around the peak at $(\pi/2,0)$. Considering the mean-field approximation, some deviations 
should be expected, and overall the agreement has to be regarded as surprisingly good. Similarly good agreement was 
previously found with the $J$-$Q$ model with spin-anisotropic interactions \cite{ma18}. The continuum along the same BZ path as in 
Fig.~\ref{fig2} is shown with a logarithmic heat map of $L=32$ data in Fig.~\ref{fig3}, along with two different calculations 
of the  upper bound of the two-spinon excitations, which we discuss next.

A $\pi$-flux phase can be realized with different gauges; the phases giving net flux $\pi$ in each  plaquette.  A $2\times 1$ 
unit cell with alternating phases $\pm \pi/2$ was used in Ref.~\cite{ma18}, with the BZ $k_x \in [0,\pi]$, $k_y \in [0,2\pi]$. The excitations
contributing to $S({\bf k},\omega)$ correspond to moving a fermion with momentum ${\bf k}_{1}$ from the filled band (creating an anti-spinon) 
with negative energy [Eq.~(\ref{epsilon_s}) with a minus sign] to the empty band at ${\bf k}_{2}$ (creating a spinon). The upper bound of
the two-spinon continuum depends on how the BZ for the individual spinons is chosen. Here, we take $|k_{x}|+|k_{y}|\leq\pi$ 
in order to preserve the symmetry between the $x$ and $y$ directions, leading to the upper bound shown with a red dashed curve in 
Fig.~\ref{fig3} (slightly different from Ref.~\cite{ma18}). This bound describes the data reasonably well, though with a glaring mismatch for $k$ 
close to $0$, where there is very little weight in a cone extending from ${\bf k}=(0,0)$. The agreement here is much improved by using  
phases $\pm \pi/4$ alternating in both directions, in which case the dispersion relation is still given by Eq.~(\ref{epsilon_s}) but now with 
the smaller BZ ${\bf k} \in [0,\pi]^2$. All levels are then doubly degenerate and the two negative bands are filled.  As shown with the white dashed 
curve in Fig.~\ref{fig3}, the upper bound close to ${\bf k}=(0,0)$ is now better compatible with the data and the agreement is also improved close 
to $(\pi,0)$. Some spectral weight should be expected also above the two-spinon bound, because of contributions from more than 
two spinons.

Comprehensive $\pi$-flux calculations with the larger BZ in Ref.~\cite{ma18} also indicate very little spectral weight 
close to $k=0$, even though the upper bound formally appears much too high. However, the weight does not have the 
characteristic $V$ shape seen in the SAC data (in Fig.~\ref{fig3} and even more clearly for the anisotropic model in Ref.~\cite{ma18}). The better 
upper bound obtained with the $2\times 2$ unit cell is analogous to the prototypical case of the two-spinon continuum in the $S=1/2$ Heisenberg
chain, where the reduced BZ ${\bf k} \in [0,\pi]^2$ for the individual spinons should be used \cite{muller81,bougourzi96,caux06}
in order for the band width to vanish as $k \to 0$, as in the Bethe Ansatz solution. If the full BZ $[0,2\pi]$ is used, the continuum instead 
has maximal support at both $k=0$ and $k=\pi$. The reduced BZ reflects the actual number of spinon degrees of freedom and can be related to semion 
statistics \cite{haldane91}. The $\pi$-flux state with ${\bf k} \in [0,\pi]^2$ has previously been used to describe deconfined spinons in 2D U($1$) 
spin liquids \cite{hermele04}, and its applicability to our case indicates a similar, but unstable spin liquid nature of the DQCP.

{\it Single-Hole Spectral Function.}---Turning to the system with an injected hole, the spectral function is given in the Lehmann representation by
\begin{equation}
A(k,\omega)=\sum_{m} |\langle m|c_{\sigma,{\bf k}}|0\rangle|^2 \delta(\omega+E_0-E_n),
\end{equation}
where $|0\rangle$ is still the ground state of the spin model but $|m\rangle$ refer to states with one hole. We use Angelucci's canonical transformation
\cite{angel95} implemented for SSE \cite{yang25b} to compute the corresponding Green's function
$G_A({\bf r},\tau)=\langle c^\dagger_{\sigma,{\bf r}}(\tau)c_{\sigma,{\bf r}}(0)\rangle$,
then use the same constrained SAC method as above for $A({\bf k},\omega)$. 

\begin{figure}[t!]
\includegraphics[width=8.4cm]{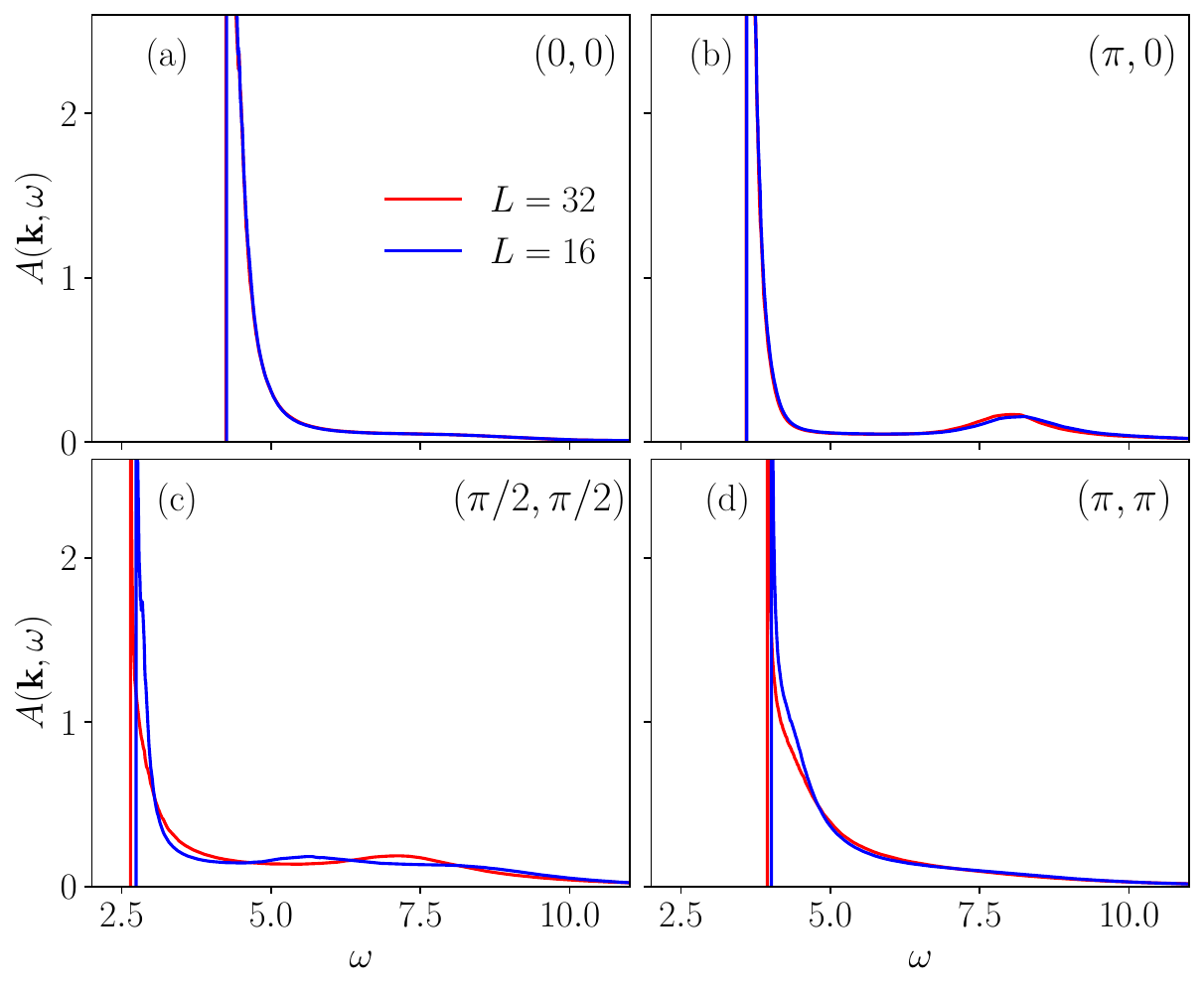}
\caption{Single-hole spectral function $A({\bf k},\omega$) of the $t$-$J$-$Q$ model with $t=J=1$, $Q=0.6667$ for lattices with $L=16$ and 
$L=32$ at four different momenta $(k_x,k_y)$.}
\label{fig4}
\end{figure}

\begin{figure}[t!]
\includegraphics[width=8.7cm]{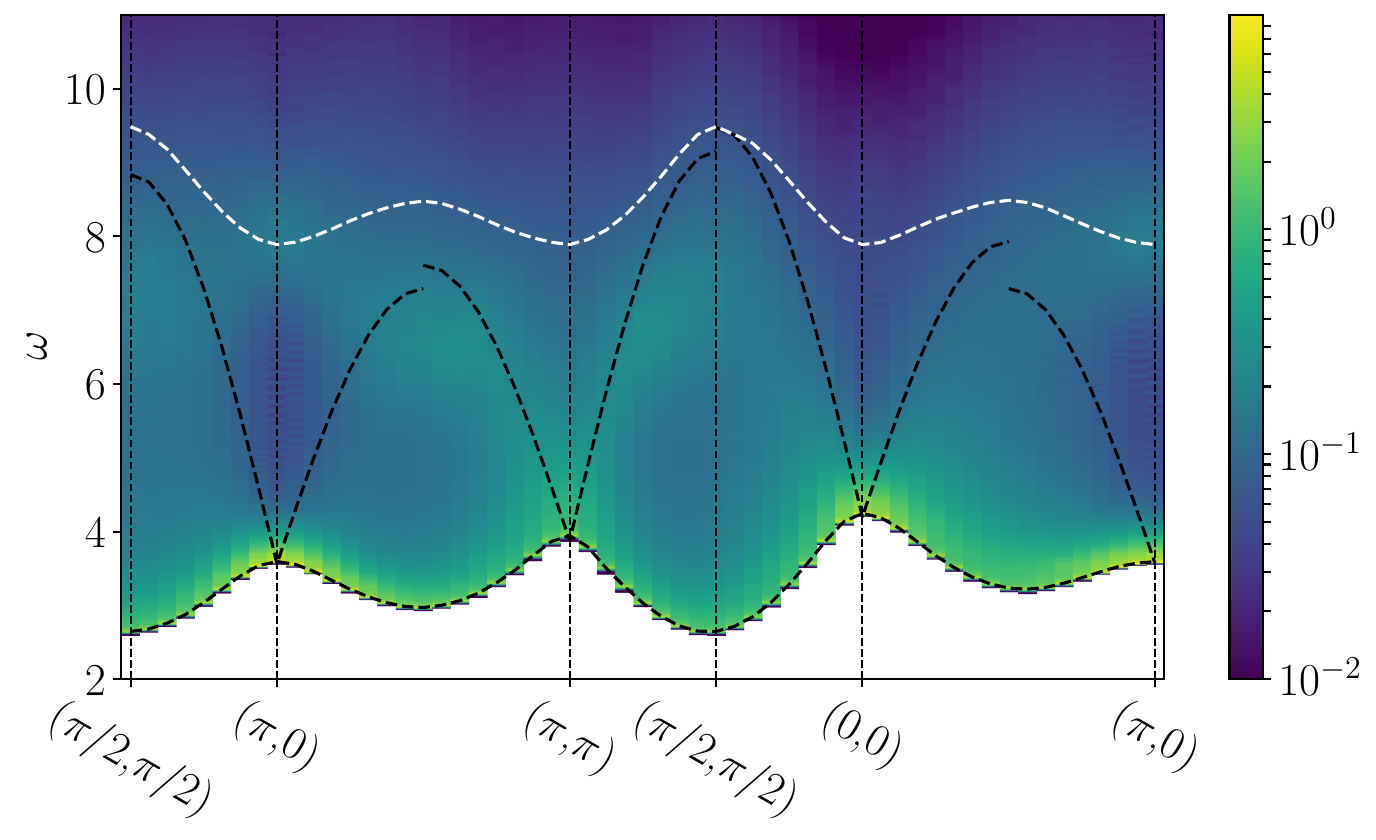}
\caption{Heat map of $A({\bf k},\omega$). The upper bound shown as the white dashed curve is obtained on the 
basis of the SAC data points for the lower edges of the two-spinon (Fig.~\ref{fig3}) and spinon-holon (this figure) continua. The black dashed lines represent
spinons added to holons with momentum fixed at one of the three peak locations. There is no spectral weight in the white area below the edge, while above
the edge any weight below $10^{-2}$ is represented as black.}
\label{fig5}
\end{figure}

Fig.~\ref{fig4} shows results at four momenta for the case $t=1$,
which we will exclusively focus on here. Finite-size effects are small in all cases. A heat map of the continuum is shown in 
Fig.~\ref{fig5} along with the upper bound constructed from the spinon dispersion (the lower edge of SAC results in the entire BZ) 
and the holon dispersion corresponding to (as it turns out) the edge of $A({\bf k},\omega)$. The edge can be reasonably well described by a tight-binding model 
with up to third-neighbor hole hopping, as shown in the Appendix. The dispersion minimum at ${\bf k}=(\pm \pi/2,\pm \pi/2)$ is similar to that of the 2D $t$-$J$ 
model \cite{schmitt88,schraiman88,dagotto90,liu92,moreo95,mis01,brunner00}, though without the symmetry with respect to a $(\pi,\pi)$ momentum 
shift expected (and confirmed \cite{yang25a}) in the presence of AFM order.

If there is spin-charge separation, the lower excitation bound at momentum ${\bf k}$ should arise from a holon at ${\bf k}_h$ and an anti-spinon at ${\bf k}_s$,
so that ${\bf k}={\bf k}_h-{\bf k}_s$ and the energy $\epsilon({\bf k}) = \epsilon_h({\bf k}_h)+\epsilon_s({\bf k}_s)$, with the spinon energy Eq.~(\ref{epsilon_s}). 
We do not a priori know the holon dispersion, but a natural assumption is that the lowest $\epsilon({\bf k})$ corresponds to a holon plus a zero-energy spinon. 
In fact, our results are consistent with the entire lower bound $\omega_{\bf k}$ arising from a holon at ${\bf k}_h={\bf k}$ and a spinon with ${\bf k}_s=0$. 
As shown in the Appendix, in order to describe $\omega_{\bf k}$ as arising from independently propagating holons and spinons, we have to shift the spinon BZ 
for Eq.~(\ref{epsilon_s}) to ${\bf k} \in [-\pi/2,\pi/2]^2$. Otherwise, the three maximums of the lower edge in Fig.~\ref{fig5} will all be of the same height, 
e.g., a holon at $(\pi,0)$ could be combined with a zero-energy anti-spinon at $(\pi,0)$, thus leading to a lower than observed peak at ${\bf k}=(0,0)$. The 
lower bound can also not be reproduced with the $\pi$-flux state with the larger BZ, for similar reasons. The holon momentum can be taken from the full BZ 
${\bf k} \in [0,2\pi]^2$. While we do not have a rigorous explanation for the spinon BZ shift, there is again an analogy with the 1D case, where the 
spin-charge separation ansatz for the single-hole dynamics of the $t$-$J$ chain is based on a different BZ for the spinon accompanying  the holon than the 
one required to reproduce the two-spinon continuum \cite{penc97,suzuura97,sorella98}. These different constraints on the partons reflect different 
mutual statistics of two spinons and a spinon and a holon.

Using the extracted spinon and holon dispersions, the upper bound of the continuum shown with the white dashed curve in Fig.~\ref{fig5} 
is obtained and matches the heat map quite well. We also show black dashed curves in the continuum that correspond to the holon momentum
${\bf k}_h$ fixed at one of its dispersion maximums ${\bf k}_{\rm max}$ and an anti-spinon with momentum $-{\bf k}_s$, 
for a total momentum ${\bf k}={\bf k}_{\rm max}+{\bf k}_s$. These curves account reasonably well for the regions of elevated spectral weight, and the 
wedges between them extending out from ${\bf k}=(\pi,0)$ and $(0,0)$ correspond to regions of low spectral weight. Broadened peaks form around these
curves when spinons are added to the holons in the neighborhood of the ${\bf k}_{\rm max}$ points. Intuitively, the elevated spectral weight in
these particular regions follows from the fact that a high holon energy corresponds to hole injection causing large perturbations of the spin background,
thus spinons radiate at large amplitude. In contrast, low-energy holons perturb the spin background only weakly and less spinons are excited.
Quantitative analytic calculations of these effects are beyond the scope of the present work.

{\it Conclusions.}---Our results strengthen the case for the $J$-$Q$ model hosting a quantum phase transition with deconfined
excitations up to large length scales, despite the ultimately weak coexisting AFM and VBS orders. The two-spinon continuum points to
semion-like statistics that require a reduced BZ in the $\pi$-flux mean-field description, which has previously been used to describe
U($1$) spin liquids phases \cite{hermele04,thomson17}. Consistency with the spinon-holon continuum (spin-charge separation in the entire BZ)
also requires this smaller spinon BZ. These results should be useful as input to field-theoretical calculations of $A({\bf k},\omega)$
and may also contribute to the resolution of the enigmatic DQCP and the nature of the proposed extended holon metal phase
\cite{kaul07}. 

Experimentally, the best candidate so far for studying fractionalization at the DQCP is SrCu$_2$(BO$_3$)$_2$ \cite{cui23}, 
though low temperature combined with high pressure and magnetic field are challenging for inelastic neutron scattering studies of 
$S({\bf k},\omega)$ \cite{zayed17} and angle-resolved photoemission spectroscopy for $A({\bf k},\omega)$. In the latter case,
charge accumulation in the insulator is also problematic \cite{hu21}, though results for spin chains \cite{kim96,kim97,kim06}
and 2D AFM systems have been obtained \cite{wells95,dama03,shen07}.

\begin{acknowledgments}
{\it Acknowledgments.}---We would like to thank Yi-Zhuang You for discussions, and Gabe Schumm and Bowen Zhao for related collaborations.
This research was supported by the Simons Foundation under Grant No.~511064. The numerical calculations were carried out on the Shared Computing
Cluster managed by Boston University’s Research Computing Services.
\end{acknowledgments}

\section*{Appendix: Holon Dispersion Relation}

\begin{figure}[t!]
\includegraphics[width=8.7cm]{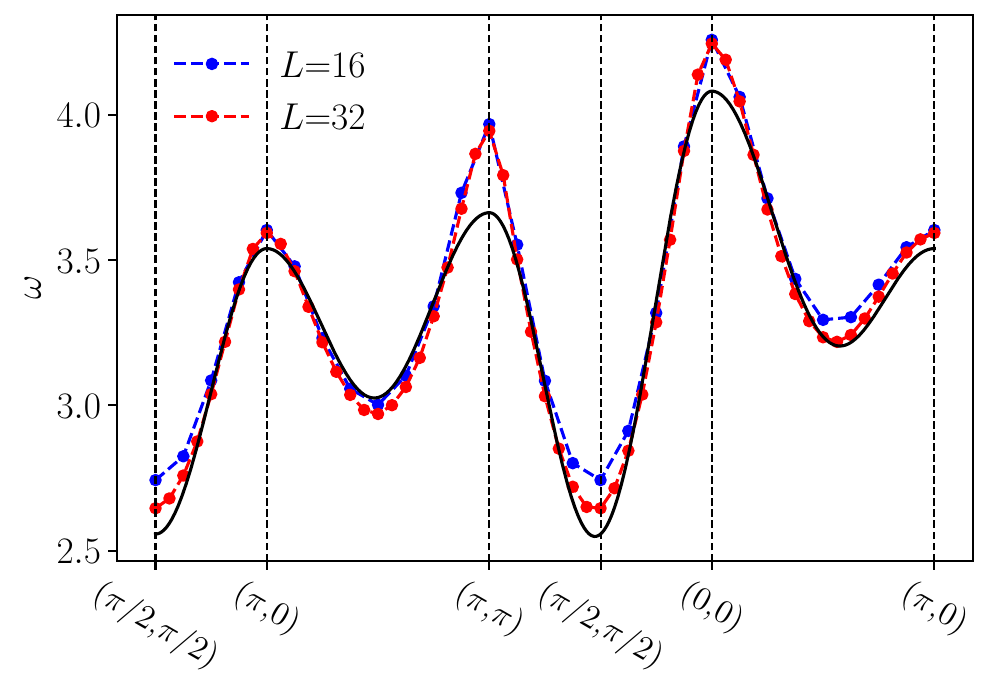}
\caption{Location of the edge in $A({\bf k},\omega$) versus the momentum along a path in the BZ, based on results such as those in Figs.~\ref{fig4} and \ref{fig5} 
for lattice sizes $L=16$ and $L=32$. The black curve shows a fit to a tight-binding model Eq.~(\ref{epsilon_h}) with 
$t_1=0.104, t_2=0.083, t_3=0.287$, $\mu=3.13$.}
\label{fig6}
\end{figure}

Figure \ref{fig6} shows the momentum dependence of the edge of $A({\bf k},\omega)$, corresponding to the single-hole dispersion relation, 
for system sizes $L=16$ and $L=32$ based on results such as those in Fig.~\ref{fig5} but on a more detailed scale. Because of some remaining 
finite-size effects, especially close to the global minimum at ${\bf k}=(\pi/2,\pi/2)$, it is not meaningful to fit all the details of the momentum dependence. 
The black solid curve shows a fit to a
tight-binding model including hopping amplitudes up to third neighbors,
\begin{eqnarray}\label{epsilon_h}
\epsilon_h({\bf k}) & = & t_1[\cos({k}_x) + \cos({k}_y)] \nonumber \\
 &  & + t_2[\cos({k}_x+{k}_y) + \cos({k}_x-{k}_y)] \nonumber \\
 &  & + t_3[\cos(2{k}_x) + \cos(2{k}_y)] + \mu,
\end{eqnarray}
with the parameters listed in the figure caption. Note that the hopping amplitudes are all positive and that the dominant $t_3$ hopping (third-neighbor,
corresponding to distance two lattice spacings) is responsible for the minimum at ${\bf k}=(\pi/2,\pi/2)$, which can be qualitatively understood as a third-order
process with a flip of a nearest-neighbor spin pair followed by two holon moves (or two holon moves followed by a spin flip). In a fully AFM ordered background,
such a hole move does not ruin the spin order, which is an important aspect of the hole dynamics in the $t$-$J$ model
\cite{schmitt88,schraiman88,dagotto90,liu92,moreo95}.
Our results show that the same dispersion minimum (now of the holon) remains also at the AFM--VBS transition. However, as we show in Ref.~\cite{yang25a}, the 
dispersion minimum shifts continuously away from $(\pm \pi/2,\pm \pi/2)$ upon moving inside the VBS phase, as previously proposed based on field-theory
calculations \cite{kaul07}.

\begin{figure}[t!]
\includegraphics[width=8.7cm]{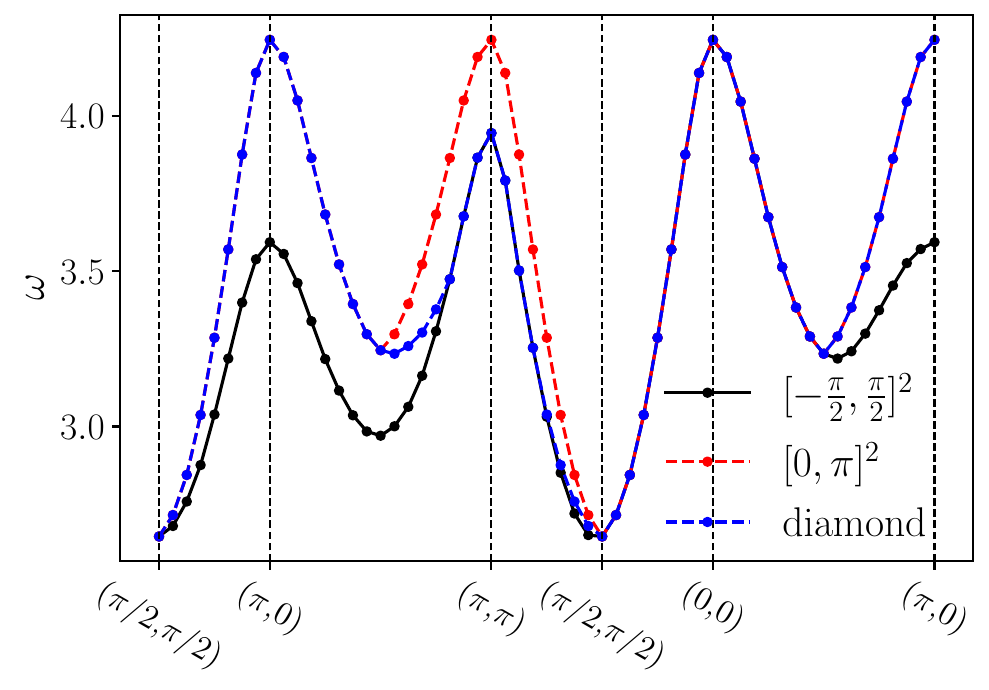}
\caption{Different extracted holon dispersion relations (partially with overlapping symbols) when using different choices of the BZ for the spinons; black circles
for ${\bf k}_s \in[-\pi/2,\pi/2]^2$, red for ${\bf k}_s \in[0,\pi]^2$, and blue for the diamond shaped BZ $|k_{sx}|+|k_{sy}|\leq\pi$. These curves also correspond 
to the edge excitation energy within the spin-charge separation scenario, with the black circles being the actual $L=32$ edge in Fig.~\ref{fig6}.}
\label{fig7}
\end{figure} 

Our conclusion that the lower bound of the hole excitation spectrum gives the holon dispersion directly is tied to our argument that the BZ of the $\pi$-flux
state should be taken as ${\bf k}_s \in [-\pi/2,\pi/2]^2$ when creating an (anti) spinon and a holon, which we explain as follows: With this BZ, the only gapless 
spinon excitations reside at ${\bf k}_s=(0,0)$, though there are still four gapless excitations of a spinon and an anti-spinon, since the fermionic 
hole (the anti-spinon) can be created in one out of the two bands with energy $\epsilon_S \le 0$ and the particle (spinon) created in one of the two
$\epsilon_S \ge 0$, Eq.~(\ref{epsilon_s}). The lowest energy of the spinon-holon pair is then with ${\bf k}_s=(0,0)$ so that ${\bf k}={\bf k}_h$.
In contrast, taking the spinon BZ as ${\bf k}_{s} \in[0,\pi]^2$, there are zero-energy spinons at $(0,0)$, $(0,\pi)$, $(\pi,0)$, and $(\pi,\pi)$. If we
then insist that the holon dispersion still corresponds to the observed edge energy, in some parts of the extended BZ it would be possible to combine a
holon and a spinon with ${\bf k}_s \not= (0,0)$ so that the total energy falls below the observed edge. To maintain the observed lowest energy for
given ${\bf k}$, the holon energy must of course be at or above the edge energy. 

Assuming spin-charge separation,
with the edge minimizing $\epsilon({\bf k}) = \epsilon_h({\bf k}-{\bf q})+\epsilon_s({\bf q})$, the holon dispersion can be extracted as the maximum of
$\epsilon_h({\bf k})=\epsilon({\bf k}+{\bf q})-\epsilon_s({\bf q})$ \cite{boudec04}. Here ${\bf q}$ is taken over all points within the
spinon BZ. In Fig.~\ref{fig7}, the red points show the holon energy obtained when this BZ is taken as ${\bf k}_s \in[0,\pi]^2$, and the blue points
similarly show the result for the twice larger diamond shaped BZ $|k_{sx}|+|k_{sy}|\leq\pi$ (which gives the upper bound of the continuum indicated 
by the red dashed line in Fig.~\ref{fig5}). These curves would also directly give the lowest total energy, corresponding again to ${\bf k}={\bf k}_h$
and ${\bf k}_s =(0,0)$, in contradiction with the observation of the actual lowest energy (and holon dispersion) shown by the black circles. The observed
lower edge is reproduced correctly if the spinons are taken from the shifted smaller BZ ${\bf k}_s \in[-\pi/2,\pi/2]^2$, then with ${\bf k}={\bf k}_h$
and ${\bf k}_s=(0,0)$ all through the extended BZ $[0,2\pi]^2$. This is then also the condition for spin-charge separation on the entire BZ.


\begin{thebibliography}{99}

\bibitem{senthil04a}  
T. Senthil, A. Vishwanath, L. Balents, S. Sachdev, and M. P. A. Fisher,
Deconfined Quantum Critical Points, Science {\bf 303}, 1490 (2004).

\bibitem{read89}
N. Read and S. Sachdev, Valence-Bond and Spin-Peierls Ground States of Low-Dimensional Quantum Antiferromagnets, 
Phys. Rev. Lett. {\bf 62}, 1694 (1989).

\bibitem{read90}
N. Read and S. Sachdev, Spin-Peierls, valence-bond solid, and N\'eel  ground states of low-dimensional quantum
antiferroma nets, Phys. Rev. B {\bf 42}, 4568 (1990).

\bibitem{murthy90}
G. Murthy and S. Sachdev, Action of hedgehog instantons in the disordered phase of the (2+1)-dimensional
CP$^{N-1}$ model, Nucl. Phys. B {\bf 344}, 557 (1990).

\bibitem{senthil04b}  
T. Senthil, L. Balents, S. Sachdev, A. Vishwanath, and M. P. A. Fisher,
Quantum criticality beyond the Landau-Ginzburg-Wilson paradigm, Phys. Rev. B {\bf 70}, 144407 (2004).

\bibitem{levin04}
M. Levin and T. Senthil,
Deconfined quantum criticality and N\'eel order via dimer disorder, Phys. Rev. B {\bf 70}, 220403 (2004).

\bibitem{hermele04}
M. Hermele, T. Senthil, M. P. A. Fisher, P. A. Lee, N. Nagaosa, and X.-G. Wen,
Stability of U($1$) spin liquids in two dimensions,
Phys. Rev. B {\bf 70}, 214437 (2004).

\bibitem{hermele05}
M. Hermele, T. Senthil, and M. P. A. Fisher, Algebraic spin liquid as the mother of many competing orders,
Phys. Rev. B {\bf 72}, 104404 (2005).

\bibitem{tanaka05}
A. Tanaka and X. Hu,
Many-Body Spin Berry Phases Emerging from the $\pi$-Flux State: Competition between Antiferromagnetism and the Valence-Bond-Solid State,
Phys. Rev. Lett. {\bf 95}, 036402 (2005).

\bibitem{senthil06}
T. Senthil and M. P. A. Fisher, Competing orders, nonlinear sigma models, and topological terms in quantum magnets,
Phys. Rev. B {\bf 74}, 064405 (2006).

\bibitem{dyer15}
E. Dyer, M. Mezei, S. S. Pufu, and S. Sachdev, Scaling dimensions of monopole operators in the CP$^{N_b-1}$
theory in $2+1$ dimensions, J. High Energy Phys. {\bf 2015}, 37 (2015); erratum {\bf 2016}, 111 (2016).

\bibitem{thomson17}
A. Thomson and S. Sachdev,
Fermionic spinon theory of square lattice spin liquids near the N\'eel state, Phys. Rev. X {\bf 8}, 011012 (2018).

\bibitem{song19}
X.-Y. Song, C. Wang, A. Vishwanath, and Y.-C. He, 
Unifying description of competing orders in two-dimensional quantum magnets,
Nat. Comm. {\bf 10}, 4254 (2019).

\bibitem{wang17}
C. Wang, A. Nahum, M. A. Metlitski, C. Xu, and T. Senthil, Deconfined Quantum Critical Points: Symmetries and Dualities, 
Phys. Rev. X {\bf 7}, 031051 (2017).

\bibitem{sandvik07}
A. W. Sandvik, Evidence for Deconfined Quantum Criticality in a Two-Dimensional Heisenberg Model with Four-Spin
Interactions, Phys. Rev. Lett. {\bf 98}, 227202 (2007).

\bibitem{melko08}
R. G. Melko and R. K. Kaul, Scaling in the Fan of an Unconventional Quantum Critical Point,
Phys. Rev. Lett. {\bf 100}, 017203 (2008).

\bibitem{kaul08}
R. K. Kaul and R. G. Melko,
Large-N estimates of universal amplitudes of the CP$^{N-1}$ theory and comparison with a $S=1/2$ square-lattice
model with competing four-spin interactions, Phys. Rev. B {\bf 78}, 014417 (2008).

\bibitem{jiang08}
F.-J. Jiang, M. Nyfeler, S. Chandrasekharan, and U.-J. Wiese, From an antiferromagnet to a valence bond solid:
Evidence for a first-order phase transition, J. Stat. Mech. {\bf 2008}, P02009 (2008).

\bibitem{kuklov08}
A. B. Kuklov, M. Matsumoto, N. V. Prokof’ev, B. V. Svistunov, and M. Troyer, 
Deconfined Criticality: Generic First-Order Transition in the SU(2) Symmetry Case,
Phys. Rev. Lett. {\bf 101}, 050405 (2008) 

\bibitem{lou09}
J. Lou, A. W. Sandvik, and N. Kawashima, Antiferromagnetic to valence-bond-solid transitions in two-dimensional
SU(N) Heisenberg models with multispin interactions, Phys. Rev. B {\bf 80}, 180414(R) (2009).

\bibitem{sandvik10a}
A. W. Sandvik, Continuous Quantum Phase Transition between an Antiferromagnet and a Valence-Bond Solid
in Two Dimensions: Evidence for Logarithmic Corrections to Scaling, Phys. Rev. Lett. {\bf 104}, 177201 (2010).

\bibitem{chen13}
K. Chen, Y. Huang, Y. Deng, A. B. Kuklov, N. V. Prokof’ev, and B. V. Svistunov,
Deconfined Criticality Flow in the Heisenberg Model with Ring-Exchange Interactions,
Phys. Rev. Lett. {\bf 110}, 185701 (2013).

\bibitem{kaul12}
R. K. Kaul and A. W. Sandvik, Lattice Model for the SU(N) N\'eel to Valence-Bond Solid Quantum Phase Transition
at Large N, Phys. Rev. Lett.{\bf  108}, 137201 (2012).

\bibitem{block13}
M. S. Block, R. G. Melko, and R. K. Kaul, Fate of CP$^{N-1}$ Fixed Points with $q$ Monopoles,
Phys. Rev. Lett. {\bf 111}, 137202 (2013).

\bibitem{nahum15a}
A. Nahum, J. T. Chalker, P. Serna, M. Ortu\~no, and A. M. Somoza,
Deconfined Quantum Criticality, Scaling Violations, and Classical Loop Models, Phys. Rev. X {\bf 5}, 041048 (2015).

\bibitem{nahum15b}
A. Nahum, P. Serna, J. T. Chalker, M. Ortu\~no,, and A. M. Somoza, Emergent SO(5) Symmetry at the N\'eell
to Valence-Bond-Solid Transition, Phys. Rev. Lett. {\bf 115}, 267203 (2015).

\bibitem{shao16}
H. Shao, W. Guo, and A. W. Sandvik, Quantum criticality with two length scales, Science {\bf 352}, 213 (2016).

\bibitem{suwa16}
H. Suwa, A. Sen, and A. W. Sandvik, Level spectroscopy in a two-dimensional quantum magnet: Linearly dispersing spinons at the deconfined 
quantum critical point, Phys. Rev. B {\bf 94}, 144416 (2016).

\bibitem{sreejith19}
G. J. Sreejith, S. Powell, and A. Nahum, Emergent SO(5) Symmetry at the Columnar Ordering Transition in the
Classical Cubic  Dimer Model, Phys. Rev. Lett. {\bf 122}, 080601 (2019).

\bibitem{zhao19}
B. Zhao, P. Weinberg, and A. W. Sandvik,
Symmetry-enhanced discontinuous phase transition in a two-dimensional quantum magnet,
Nat. Phys. {\bf 15}, 678 (2019).

\bibitem{serna19}
P. Serna and A. Nahum,
Emergence and spontaneous breaking of approximate O($4$) symmetry at a weakly first-order deconfined phase
transition, Phys. Rev. B {\bf 99}, 195110 (2019).

\bibitem{zhao20}
B. Zhao, J. Takahashi, and A. W. Sandvik, Multicritical Deconfined Quantum Criticality and Lifshitz Point
of a Helical Valence-Bond Phase, Phys. Rev. Lett. {\bf 125}, 257204 (2020).

\bibitem{sandvik20}
A. W. Sandvik and B. Zhao, Consistent scaling exponents at the deconfined quantum-critical point, Chin. Phys. Lett. {\bf 37}, 057502 (2020).

\bibitem{takahashi20}
J. Takahashi and A. W. Sandvik, Valence-bond solids, vestigial order, and emergent SO(5) symmetry in a
two-dimensional quantum magnet, Phys. Rev. Research {\bf 2}, 033459 (2020).

\bibitem{wang21}
Z. Wang, M. P. Zaletel, R. S. K. Mong, and F. F. Assaad, Phases of the (2+1) Dimensional SO(5) Nonlinear Sigma Model with 
Topological Term, Phys. Rev. Lett. {\bf 126}, 045701 (2021).

\bibitem{zhao22}
J. Zhao, Y.-C. Wang, Z. Yan, M. Cheng, and Z. Y. Meng,
Scaling of Entanglement Entropy at Deconfined Quantum Criticality,
Phys. Rev. Lett. {\bf 128}, 010601 (2022).

\bibitem{sato23}
T. Sato, Z. Wang, Y. Liu, D. Hou, M. Hohenadler, W. Guo, and F. F. Assaad,
Simulation of fermionic and bosonic critical points with emergent SO(5) symmetry,
Phys. Rev. B {\bf 108}, L121111 (2023).

\bibitem{takahashi24}
J. Takahashi, H. Shao, B. Zhao, W. Guo, and A. W. Sandvik, SO(5) multicriticality in two-dimensional quantum magnets
arXiv2405.06607.

\bibitem{demidio24}
J. D'Emidio and A. W. Sandvik, Entanglement entropy and deconfined criticality: emergent SO(5) symmetry and proper lattice bipartition,
Phys. Rev. Lett. {\bf 133}, 166702 (2024).

\bibitem{song25}
M. Song, J. Zhao, M. Cheng, C. Xu, M. Scherer, L. Janssen, and Z. Y. Meng,
Evolution of entanglement entropy at SU(N) deconfined quantum critical points, Sci. Adv. {\bf 11}, eadr0634 (2025).

\bibitem{nakayama16}
Y. Nakayama and T. Ohtsuki, Necessary Condition for Emergent Symmetry from the Conformal Bootstrap
Phys. Rev. Lett. {\bf 117}, 131601 (2016).

\bibitem{li19}
Z. Li, Bootstrapping conformal QED3 and deconfined quantum critical point, arXiv:1812.09281.
 
\bibitem{poland19}
D. Poland, S. Rychkov, and A. Vichi,
The Conformal Bootstrap: Theory, Numerical Techniques, and Applications,
Rev. Mod. Phys. {\bf 91}, 15002 (2019).

\bibitem{chester24}
S. M. Chester and N. Su,  Bootstrapping Deconfined Quantum Tricriticality, 
Phys. Rev. Lett. {\bf 132}, 111601 (2024).

\bibitem{zhou24}
Z. Zhou, L. Hu, W. Zhu, Y.-C. He,
The SO(5) Deconfined Phase Transition under the Fuzzy Sphere Microscope:
Approximate Conformal Symmetry, Pseudo-Criticality, and Operator Spectrum, Phys. Rev. X {\bf 14}, 021044 (2024).

\bibitem{chen24}
B.-B. Chen, X. Zhang, Y. Wang, K. Sun, and Z. Y. Meng,
Phases of (2+1)D SO(5) non-linear sigma model with a topological term on a sphere: multicritical point and disorder phase, Phys. Rev. Lett. {\bf 132}, 246503 (2024).

\bibitem{gorbenko18a}
V. Gorbenko, S. Rychkov, and B. Zan, Walking, weak first-order transitions, and complex CFTs, J. High Energy Phys. {\bf 10},  108 (2018).

\bibitem{gorbenko18b}
V. Gorbenko, S. Rychkov, and B. Zan, Walking, Weak first-order transitions, and Complex CFTs II. Two-dimensional Potts model at 
$Q > 4$, SciPost Phys. {\bf 5}, 50 (2018).

\bibitem{ma20}
R. Ma and C. Wang, A theory of deconfined pseudocriticality, Phys. Rev. B {\bf 102}, 020407(R) (2020).

\bibitem{nahum20}
A. Nahum, Note on Wess-Zumino-Witten models and quasi-universality in 2+1 dimensions, Phys. Rev. B {\bf 102}, 201116(R) (2020).

\bibitem{he21}
Y.-C. He, J. Rong, and N. Su, Non-Wilson-Fisher kinks of O(N) numerical bootstrap: from the deconfined phase transition to a putative new family 
of CFTs, SciPost Phys. {\bf 10}, 115 (2021).

\bibitem{schackleton21}
H. Shackleton, A. Thomson, and S. Sachdev, Deconfined criticality and a gapless Z$_2$ spin liquid in the square
lattice  antiferromagnet, Phys. Rev. B {\bf 104}, 045110 (2021).

\bibitem{yang22}
J. Yang, A. W. Sandvik, and L. Wang, 
Quantum criticality and spin liquid phase in the Shastry-Sutherland model,
Phys. Rev. B {\bf 105}, L060409 (2022). 

\bibitem{liu22}
W.-Y. Liu, J. Hasik, S.-S. Gong, D. Poilblanc, W.-Q. Chen, and Z.-C. Gu,
Emergence of Gapless Quantum Spin Liquid from Deconfined Quantum Critical Point,
Phys. Rev. X {\bf 12}, 031039 (2022).

\bibitem{liu24}
W.-Y. Liu, D. Poilblanc, S.-S. Gong, W.-Q. Chen, and Z.-C. Gu,
Tensor network study of the spin-$\frac{1}{2}$ square-lattice $J_1$-$J_2$-$J_3$ model: Incommensurate spiral order, 
mixed valence-bond solids, and multicritical points,
Phys. Rev. B {\bf 109}, 235116 (2024).

\bibitem{ma18}
N. Ma, G.-Y. Sun, Y.-Z. You, C. Xu, A. Vishwanath, A. W. Sandvik, and Z. Y. Meng,
Dynamical signature of fractionalization at a deconfined quantum critical point
Phys. Rev. B {\bf 98}, 174421 (2018).

\bibitem{kaul07}
R. K. Kaul, A. Kolezhuk, M. Levin, S. Sachdev, and T. Senthil,
Hole dynamics in an antiferromagnet across a deconfined quantum critical point, Phys. Rev. B {\bf 75}, 235122 (2007).

\bibitem{loh90}
E. Y. Loh, Jr., J. E. Gubernatis, R. T. Scalettar, S. R. White, D. J. Scalapino, and R. L. Sugar,
Sign problem in the numerical simulation of many-electron systems,
Phys. Rev. B {\bf 41}, 9301 (1990).

\bibitem{sandvik99}
A. W. Sandvik, Stochastic series expansion method with operator-loop update,
Phys. Rev. B {\bf 59}, R14157 (1999).  

\bibitem{sandvik10}
A. W. Sandvik, Computational Studies of Quantum Spin Systems, AIP Conf. Proc. {\bf 1297}, 135 (2010).

\bibitem{shao23}
H. Shao and A. W. Sandvik, Progress on stochastic analytic continuation of quantum Monte Carlo data,
Phys. Rep. {\bf 1003}, 1 (2023).

\bibitem{yang25a}
S. Yang and A. W. Sandvik (in preparation).

\bibitem{angel95}
A. Angelucci, Effective lattice actions for correlated electrons, Phys. Rev. B {\bf 51}, 11580 (1995).

\bibitem{brunner00}
M. Brunner, F. F. Assaad, and A. Muramatsu, Single-hole dynamics in the $t$-$J$ model on a square lattice,
Phys. Rev. B {\bf 62}, 15480 (2000).

\bibitem{yang25b}
S. Yang, G. Schumm, B. Zhao, and A. W. Sandvik,
Single-hole spectral functions in 1D quantum magnets with different ground states,
arXv:2511.20447.

\bibitem{yang25c}
S. Yang, G. Schumm, and A. W. Sandvik, Dynamic structure factor of a spin-1/2 Heisenberg chain with long-range interactions,
Phys. Rev. B {\bf 111}, 224404 (2025).

\bibitem{muller81}
Quantum spin dynamics of the antiferromagnetic linear chain in zero and nonzero magnetic field
G. M\"uller, H. Thomas, H. Beck, and J. C. Bonner, Phys. Rev. B {\bf 24}, 1429 (1981)

\bibitem{bougourzi96}
A. H. Bougourzi, M. Couture, and M. Kacir,
Exact two-spinon dynamical correlation function of the one-dimensional Heisenberg model,
Phys. Rev. B {\bf 54} R12669 (1996).

\bibitem{caux06}
J.-S. Caux and R. Hagemans, The four-spinon dynamical structure factor of the Heisenberg chain,
J. Stat. Mech. {\bf 2006}, 12013 (2006).

\bibitem{haldane91}
F. D. M. Haldane, "Fractional Statistics" in Arbitrary Dimensions: A Generalization of the Pauli Principle,
Phys. Rev. Lett. {\bf 67}, 937 (1991).

\bibitem{penc97}
K. Penc, K. Hallberg, F. Mila, and H. Shiba,
Spectral functions of the one-dimensional Hubbard model in the $U \to +\infty$ limit: How to use the factorized wave function
Phys. Rev. B {\bf 55} 15475 (1997).

\bibitem{suzuura97}
H. Suzuura and N. Nagaosa, Spin-charge separation in angle-resolved photoemission spectra,
Phys. Rev. B {\bf 56}, 3548 (1997).

\bibitem{sorella98}
S. Sorella and A. Parola, Theory of hole propagation in one-dimensional insulators and superconductors,
Phys. Rev. B {\bf 57}, 6444 (1998).
  
\bibitem{schmitt88}
S. Schmitt-Rink, C. M. Varma, and A. E. Ruckenstein, Spectral Function of Holes in a Quantum Antiferromagnet,
Phys. Rev. Lett. {\bf 60}, 2793 (1988).

\bibitem{schraiman88}
B. Shraiman and E. Siggia, Mobile Vacancies in a Quantum Heisenberg Antiferromagnet
Phys. Rev. Lett. {\bf 61}, 467 (1988).

\bibitem{dagotto90}
E. Dagotto, R. Joynt, A. Moreo, S. Bacci and E. Gagliano,
Strongly correlated electronic systems with one hole: Dynamical properties,
Phys. Rev. B {\bf 41} 9049 (1990).

\bibitem{liu92}
Z. Liu and E. Manousakis, Dynamical properties of a hole in a Heisenberg antiferromagnet,
Phys. Rev. B {\bf 45}, 2425 (1992).

\bibitem{moreo95}
A. Moreo, S. Haas, A. W. Sandvik, and E. Dagotto
Quasiparticle dispersion of the t-J and Hubbard models,
Phys. Rev. B {\bf 51}, 12045(R) (1995).

\bibitem{mis01}
A. S. Mishchenko, N. V. Prokof’ev, and B. V. Svistunov, Single-hole spectral function and spin-charge separation
in the $t$-$J$ model, Phys. Rev. B {\bf 64}, 033101 (2001).

\bibitem{cui23}
Y. Cui, L. Liu, H. Lin, K.-H. Wu, W. Hong, X. Liu, C. Li, Z. Hu, N. Xi, S. Li, R. Yu, A. W. Sandvik, and W. Yu , 
Proximate deconfined quantum critical point in SrCu$_2$(BO$_3$)$_2$, Science {\bf 380}, 1179 (2023).

\bibitem{zayed17}
M. E. Zayed, Ch. R{\"u}egg, J. Larrea J. , A. M. L{\"a}uchli, C. Panagopoulos, S. S. Saxena, M. Ellerby, D. F. McMorrow, Th. Str{\"a}ssle, S. Klotz, G. Hamel, R. A. Sadykov, V. Pomjakushin, M. Boehm, M. Jim{\'e}nez--Ruiz, A. Schneidewind, E. Pomjakushina, M. Stingaciu, K. Conder, and H. M. R{\o}nnow, 
4-spin plaquette singlet state in the Shastry-Sutherland compound SrCu$_2$(BO$_3$)$_2$, Nat. Phys. {\bf 13}, 962 (2017).

\bibitem{hu21}
C. Hu, J. Zhao, Q. Gao, H. Yan, H. Rong, J. Huang, J. Liu, Y. Cai, C. Li, H. Chen, L. Zhao, G. Liu, C. Jin, Z. Xu, T. Xiang, and X. J. Zhou, 
Momentum-resolved visualization of electronic evolution in doping a Mott insulator,
Nat. Comm. {\bf 12}, 1356 (2021). 

\bibitem{kim96}
C. Kim, A. Y. Matsuura, Z.-X. Shen, N. Motoyama, H. Eisaki, S. Uchida, T. Tohyama, and S. Maekawa, 
Observation of Spin-Charge Separation in One-Dimensional SrCuO$_2$, 
Phys. Rev. Lett. {\bf 77}, 4054 (1996).

\bibitem{kim97}
C. Kim, Z.-X. Shen, N. Motoyama, H. Eisaki, S. Uchida, T. Tohyama, and S. Maekawa, 
Separation of spin and charge excitations in one-dimensional SrCuO$_2$, 
Phys. Rev. B {\bf 56}, 15589 (1997).

\bibitem{kim06}
B. J. Kim, H. Koh, E. Rotenberg, S.-J. Oh, H. Eisaki, N. Motoyama, S. Uchida, T. Tohyama, S. Maekawa, Z.-X. Shen, and C. Kim, 
Distinct spinon and holon dispersions in photoemission spectral functions from one-dimensional SrCuO$_2$, 
Nat. Phys. {\bf 2}, 397 (2006).

\bibitem{wells95}
B. O. Wells, Z.-X. Shen, A. Y. Matsuura, D. M. King, M. A. Kastner, M. Greven, and R. J. Birgeneau,
E versus k Relations and Many Body Effects in the Model Insulating Copper Oxide Sr$_2$CuO$_2$Cl$_2$,
Phys. Rev. Lett. {\bf 74}, 964 (1995).

\bibitem{dama03}
A. Damascelli, Z. Hussain, and Z.-X. Shen,
Angle-resolved photoemission studies of the cuprate superconductors,
Rev. Mod. Phys. {\bf 75}, 473 (2003).

\bibitem{shen07}
K. M. Shen, F. Ronning, W. Meevasana, D. H. Lu, N. J. C. Ingle, F. Baumberger, W. S. Lee, L. L. Miller,
Y. Kohsaka, M. Azuma, M. Takano, H. Takagi, and Z.-X. Shen, Angle-resolved photoemission studies of lattice polaron formation in the cuprate Ca$_2$CuO$_2$Cl$_2$,
Phys. Rev. B {\bf 75}, 075115 (2007).

\bibitem{boudec04}
J. Y. Le Boudec and P. Thiran, 
Min-plus and Max-plus System Theory Applied to Communication Networks,
LNCIS {\bf 294}, 7 (2004).
  
\end{thebibliography}
\end{document}